\pdfoutput=1
\documentclass{ecai}
\usepackage{graphicx}
\usepackage{latexsym}

\ecaisubmission   
              
\usepackage{xspace}
\newcommand\IPAs[0]{\textsc{IPAs}\xspace}
\newcommand\IPA[0]{\textsc{IPA}\xspace}
\newcommand\AST[0]{\textsc{AST}\xspace}
\newcommand\ASTs[0]{\textsc{ASTs}\xspace}
\newcommand\GNN[0]{\textsc{GNN}\xspace}
\newcommand\GNNs[0]{\textsc{GNNs}\xspace}
\newcommand{\benchmark}[0]{\textsc{C-Pack-IPAs}\xspace}
\newcommand{\multIPAs}[0]{\textsc{Mult\-IPAs}\xspace}
\newcommand\Verifix[0]{\textsc{Ve\-ri\-fix}\xspace}
\newcommand\Clara[0]{\textsc{Cla\-ra}\xspace}

\usepackage{listings}
\usepackage{color}
\usepackage{booktabs}
\usepackage{float}
\usepackage{amsmath}
\usepackage{subcaption}
\usepackage{hyperref}
\usepackage{adjustbox}

\makeatletter
\newcommand*\mysize{%
  \@setfontsize\mysize{9.5}{9.5}%
}
\makeatother

\usepackage[frozencache,cachedir=.]{minted}

\setminted{fontsize=\mysize}

\newenvironment{code}{\captionsetup{type=listing, labelfont=bf,justification=raggedright, singlelinecheck=false}}{}

\usepackage{tikz}
\usepackage{tikz-cd}
\usetikzlibrary{arrows,automata,backgrounds,calc,chains,external,fit,graphs,patterns,positioning,shapes,shapes.geometric}

\begin{document}

\begin{frontmatter}

\title{Graph Neural Networks For Mapping Variables Between Programs - Extended Version}

\author[A]{\fnms{Pedro}~\snm{Orvalho}
\orcid{0000-0002-7407-5967}
\thanks{Corresponding Author: pmorvalho@inesc-id.pt. This work was done while this author was visiting CIIRC, CTU in Prague.}
}
\author[B,C]{\fnms{Jelle}~\snm{Piepenbrock}
\orcid{0000-0002-8385-9157}
}
\author[C]{\fnms{Mikoláš}~\snm{Janota}
\orcid{0000-0003-3487-784X}
} 
\author[A]{\fnms{Vasco}~\snm{Manquinho}
\orcid{0000-0002-4205-2189}
} 

\address[A]{INESC-ID, IST, Universidade de Lisboa}
\address[B]{Radboud University Nijmegen, The Netherlands}
\address[C]{Czech Technical University in Prague, Czechia}

\begin{abstract}
Automated program analysis is a pivotal research domain in many areas of Computer Science --- Formal Methods and Artificial Intelligence, in particular.
Due to the undecidability of the problem of program equivalence, comparing two programs is highly challenging. Typically, in order to compare two programs, a relation between both programs' sets of variables is required. Thus, mapping variables between two programs is useful for a panoply of tasks such as program equivalence, program analysis, program repair, and clone detection.
In this work, we propose using graph neural networks (\GNNs) to map the set of variables between two programs based on both programs' abstract syntax trees (\ASTs). To demonstrate the strength of variable mappings, we present three use-cases of these mappings on the task of \emph{program repair} to fix well-studied and recurrent bugs among novice programmers in introductory programming assignments (\IPAs).
Experimental results on a dataset of 4166 pairs of incorrect/correct programs show that our approach correctly maps 83\% of the evaluation dataset. Moreover, our experiments show that the current state-of-the-art on program repair, greatly dependent on the programs' structure, can only repair about 72\% of the incorrect programs. In contrast, our approach, which is solely based on variable mappings, can repair around 88.5\%.
\end{abstract}

\end{frontmatter}

\section{Introduction}
\label{sect:introduction}

The problem of program equivalence, i.e., deciding if two programs are equivalent, is undecidable~\cite{rice1953classes,pldi19-semantic-equivalence-checking}. On that account, the problem of repairing an incorrect program based on a correct implementation is very challenging.
In order to compare both programs, i.e., the correct and the faulty implementation, program repair tools first need to find a relation between both programs' sets of variables. Besides \emph{program repair}~\cite{verifix}, the task of mapping variables between programs is also important for \emph{program analysis}~\cite{oopsla20-learning-prog-embeddings-ginn}, \emph{program equivalence}~\cite{overcode15}, \emph{program clustering}~\cite{invAASTCluster-arxiv22,iclr18-neural-program-embeddings}, \emph{program synthesis}~\cite{orvalho-vldb20}, \emph{clone detection}~\cite{deckard}, and \emph{plagiarism detection}~\cite{moss}.

Due to a large number of student enrollments every year in programming courses, providing feedback to novice students in \emph{introductory programming assignments} (\IPAs) requires substantial time and effort by the faculty~\cite{drRepair}. 
Hence, there is an increasing need for systems capable of providing automated, comprehensive, and personalized feedback to students in 
programming assignments~\cite{deepfix,clara,rlassist,verifix}.
\emph{Semantic program repair} has become crucial to provide feedback to each novice programmer by checking their \IPAs submissions using a pre-defined test suite.
Semantic program repair frameworks use a correct implementation, provided by the lecturer or submitted by a previously enrolled student, to repair a new incorrect student's submission.
However, the current state-of-the-art tools on semantic program repair~\cite{clara,verifix} for \IPAs have two main drawbacks: (1) require a perfect match between the control flow graphs (loops, functions) of both programs, the correct and the incorrect one; and (2) require a bijective relation between both programs' sets of variables. Hence, if one of these requirements is not satisfied, then, these tools cannot fix the incorrect program with the correct one.

For example, consider the two programs presented in Figure~\ref{fig:progs-motivation}. These programs are students' submissions for the \IPA of printing all the natural numbers from $1$ to a given number $n$. The program in Listing~\ref{prog:cor_prog} is a semantically correct implementation that uses a for-loop to iterate all the natural numbers until $n$. The program in Listing~\ref{prog::inc_prog} uses a while-loop and an auxiliary function. This program is semantically incorrect since the student forgot to initialize the variable $j$, a frequent bug among novice programmers called \emph{missing expression/assignment}~\cite{codeFlaws-dataset}.
However, in this case, state-of-the-art program repair tools~\cite{clara,verifix} cannot fix the buggy program, since the control flow graphs do not match either due to using different loops (for-loop vs.\ while-loop) or due to the use of an auxiliary function. 
Thus, these program repair tools cannot leverage on the correct implementation in Listing~\ref{prog:cor_prog} to repair the faulty program in Listing~\ref{prog::inc_prog}.

\begin{figure*}[t!]
\centering
\begin{adjustbox}{valign=t,minipage=0.49\linewidth}
\begin{code}
\captionof{listing}{A semantically correct student's implementation.\\
}
\label{prog:cor_prog}
\vspace{0.15in}
\begin{minted}[escapeinside=||,tabsize=4,obeytabs,xleftmargin=20pt,linenos,]{C}
int main(){
    int n, i;
    scanf("%d", &n);
    for(i = 1; i <= n; i++){
        printf("%d\n", i);
    }
    return 0;
}
\end{minted}
\end{code}
\end{adjustbox}
\hfill
\begin{adjustbox}{valign=t,minipage=0.49\linewidth}
\begin{code}
\captionof{listing}{A semantically incorrect student's implementation since the variable \texttt{j} in the \texttt{main} function is not initialized.}
\label{prog::inc_prog}
\begin{minted}[escapeinside=||,tabsize=4,obeytabs,xleftmargin=20pt,linenos]{C}
void loop(int j, int l){
  while (l >= j){
    printf("%d\n", j);
    ++j;
  }
}
int main(){
  int j, l;
  scanf("%d", &l);
  loop(j, l);
  return 0;
}
\end{minted}
\end{code}
\end{adjustbox}
\caption{Two implementations for the \IPA of printing all the natural numbers from $1$ to a given number $n$. The program in Listing~\ref{prog::inc_prog} is semantically incorrect since the variable \texttt{j}, which is the variable being used to iterate over all the natural numbers until the number \texttt{l}, is not being initialized, i.e., the program has a bug of \emph{missing expression}. The mapping between these programs' sets of variables is \{\texttt{n} : \texttt{l};  \texttt{i} : \texttt{j}\}.}
\label{fig:progs-motivation}
\end{figure*}
To overcome these limitations, in this paper, we propose a novel graph program representation based on the structural information of the \emph{abstract syntax trees (\ASTs)} of imperative programs to learn how to map the set of variables between two programs using \emph{graph neural networks (\GNNs)}. Additionally, we present use-cases of program repair where these variable mappings can be applied to repair common bugs in incorrect students' programs that previous tools are not always capable of handling.
For example, consider again the two programs presented in Figure~\ref{fig:progs-motivation}. Note that having a mapping between both programs' variables (e.g. \{\texttt{n} : \texttt{l};  \texttt{i} : \texttt{j}\}) lets us reason about, on the level of expressions, which program fixes one can perform on the faulty program in Listing~\ref{prog::inc_prog}. In this case, when comparing variable \texttt{i} with variable \texttt{j} one would find the \emph{missing assignment} i.e., \texttt{j = 1}. 

Another useful application for mapping variables between different programs is fault localization. There is a body of research on fault localization~\cite{cav11-bugassist,fmcad11-fault-loc, icfem14-fault-loc-formula, jip16-fault-loc-formula-multi-faults}, that requires the usage of assertions in order to verify programs. Variable mappings can be helpful in sharing these assertions among different programs.
Additionally, several program repair techniques (e.g., \textsc{SearchRepair}~\cite{searchRepair}, \Clara~\cite{clara}) enumerate all possible mappings between two programs' variables during the search for possible fixes, using a correct program~\cite{clara} or code snippets from a database~\cite{searchRepair}. Thus, variable mappings can drastically reduce the search space, by pruning all the other solutions that use a different mapping.

In programming courses, unlike in production code, typically, there is a reference implementation for each programming exercise. This comes with the challenge of comparing different names and structures between the reference implementation and a student's program. To deal with this challenging task, we propose to map variables between programs using \GNNs.
Therefore, we explore three tasks to illustrate the advantages of using variable mappings to repair some frequent bugs without considering the incorrect/correct programs' control flow graphs. Hence, we propose to use our variable mappings to fix bugs of: \textit{wrong comparison operator}, \textit{variable misuse}, and \textit{missing expression}. These bugs are recurrent among novice programmers~\cite{codeFlaws-dataset} and have been studied by prior work in the field of automated program repair~\cite{iclr18-learning-to-represent-progs-as-graphs,oopsla18-deepBugs,iclr19-learning-localize-and-repair,neurips21-bug-detection-repair}.

Experiments on 4166 pairs of incorrect/correct programs show that our \GNN model correctly maps 83\% of the evaluation dataset. Furthermore, we also show that previous approaches can only repair about 72\% of the dataset, mainly due to control flow mismatches. On the other hand, our approach, solely based on variable mappings, can fix 88.5\%.

The main contributions of this work are:
\begin{itemize}
    \item A novel graph program representation that is agnostic to the names of the variables and for each variable in the program contains a representative variable node that is connected to all the variable's occurrences; 
    \item We propose to use \GNNs for mapping variables between programs based on our program representation, ignoring the variables' identifiers;
    \break
    \break
    \item Our \GNN model and the dataset used for this work's training and evaluation, will be made open-source and publicly available on GitHub:~\href{https://github.com/pmorvalho/ecai23-GNNs-for-mapping-variables-between-programs}{https://github.com/pmorvalho/ecai23-GNNs-for-mapping-variables-between-programs}.
\end{itemize}

The structure of the remainder of this paper is as follows.
First, Section~\ref{sec:prog-repr} presents our graph program representations. Next, Section~\ref{sec:gnns} describes the \GNNs used in this work. Section~\ref{sec:use-cases} introduces typical program repair tasks, as well as our program repair approach using variable mappings. Section~\ref{sec:results} presents the experimental evaluation where we show the effectiveness of using \GNNs to produce correct variable mappings between programs. Additionally, we compare our program repair approach based on the variable mappings generated by the \GNN with state-of-the-art program repair tools. Finally, Section~\ref{sec:related-work} describes related work, and the paper concludes in Section~\ref{sec:conclusions}.

\tikzstyle{legend} = [circle, 
    text width=12em, 
    text centered, minimum height=2em]
\tikzstyle{empt} = [circle, 
    text width=2em, 
    text centered, minimum height=2em]
\tikzstyle{statement} = [circle, draw, 
    text width=4.5em, 
    text centered, minimum height=2em]
\tikzstyle{var} = [draw, circle,fill=green!95!black, 
    text width=3em, text centered, minimum height=2em]

\begin{figure*}[t!]
\centering
\begin{minipage}[t!]{0.49\linewidth}
\begin{code}
\captionof{listing}{Small example of a C code block with an expression.
}
\label{code:simple-expr}
\begin{minted}[escapeinside=||,tabsize=4,obeytabs,xleftmargin=20pt,linenos]{C}
{ // a and b are ints
  a = a - b;
}
\end{minted}
\end{code}

\begin{subfigure}[t!]{1\linewidth}
         \centering
          \scalebox{0.6}{\begin{tikzpicture}[
              level distance=19
              mm,
              level 1/.style={sibling distance=8em},
              every node/.style = {shape=circle, draw, align=center, top color=white, bottom color=white!20}]
              \node[label={[label distance=0cm]0:}][draw] at (0, 0) {\Large block}
              [-Stealth] child { node[label={[label distance=0cm]0:}] {\Large assign}
                child { node[label={[label distance=0cm]0:{\Large a}}] {\Large ID}}
                child { node[label={[label distance=0cm]0:}] {\Large expr}
                    child { node[label={[label distance=0cm]0:{\Large a}}] {\Large ID}}
                    child { node[label={[label distance=0cm]0:{\Large $-$}}] {\Large op}}
                    child { node[label={[label distance=0cm]0:{\Large b}}] {\Large ID}}}
              };
          \end{tikzpicture}}
\caption{Part of the \AST representation of Listing~\ref{code:simple-expr}.}
\label{fig:AST}
\end{subfigure}
\end{minipage}
\hfill
\begin{subfigure}[t!]{0.5\textwidth}
\centering
\scalebox{0.45}{
\begin{tikzpicture}
\graph{
n1[statement, as={\huge block}];
n2[statement, below of=n1, node distance=1.5cm,as={\huge assign}];
n1 <->[line width=1.5pt] n2;
n3a[statement, below of=n2, xshift=-2cm, node distance=0.8cm,as={\huge ID}];
n3b[statement, below of=n2, right of=n3a, node distance=1.5cm, xshift=4cm, yshift=4.5cm, as={\huge expr}];
n2 <->[line width=1.5pt] n3a;
n2 <->[line width=1.5pt] n3b;
n3a ->[line width=1.5pt, dashed, red] n3b;
n4a[var, below of=n3a, node distance=1.2cm,  yshift=-0.5cm, as=\texttt{\huge a}];
n3a <->[line width=1.5pt, dashed, blue] n4a;
n4b[statement, below of=n3b, node distance=1.1cm, yshift=3cm, xshift=-3cm, as={\huge ID}];
n4c[statement, below of=n3b, node distance=1.1cm, yshift=4cm, xshift=0cm, as={\huge $-$}];
n4d[statement, below of=n3b, node distance=1.1cm, yshift=5cm, xshift=3cm, as={\huge ID}];
n3b <->[line width=1.5pt, line width=1.5pt] n4b;
n4b ->[line width=1.5pt, dashed, red] n4c;
n4b <->[line width=1.5pt, dashed, green!80!black] n4a;
n3a ->[line width=1.5pt, dashed, yellow!95!black] n4b;
n3b <->[line width=1.5pt] n4c;
n4c ->[line width=1.5pt, dashed, red] n4d;
n3b <->[line width=1.5pt] n4d;
n5[var, below of=n4d, node distance=1.5cm, yshift=7cm, as=\texttt{\huge b}];
n5 <->[line width=1.5pt, dashed, green!90!black] n4d;
l1[legend, xshift=-10cm, yshift=7cm, as={\huge AST}]; 
l1b[empt, right of=l1, xshift=3cm, yshift=10cm, as={}];
l1 <->[line width=1.5pt] l1b;
l0[legend, above of=l1, yshift=11cm, as={\huge Types of edges:}];
l2[legend, below of = l1, yshift=12cm, as={\huge Sibling }]; 
l2b[empt, right of= l2, xshift=3cm, yshift=13cm, as={}];
l2 ->[line width=1.5pt, dashed, red] l2b;
l3[legend, below of = l2, yshift=14cm, as={\huge Write}]; 
l3b[empt, right of= l3, xshift=3cm, yshift=15cm, as={}];
l3 <->[line width=1.5pt, dashed, blue] l3b;
l4[legend, below of = l3, yshift=16cm, as={\huge Read}]; 
l4b[empt, right of= l4, xshift=3cm, yshift=17cm, as={}];
l4 <->[line width=1.5pt, dashed, green] l4b;
l5[legend, below of = l4, yshift=18cm, as={\huge Chronological}]; 
l5b[empt, right of= l5, xshift=3cm, yshift=19cm, as={}];
l5 ->[line width=1.5pt, dashed, yellow] l5b;
l6[legend, below of = l5, yshift=19cm, as={\huge Variable Node}]; 
l6b[var, right of= l6, xshift=2cm, yshift=21cm, as={}];
};
\end{tikzpicture}}
\caption{Our program representation for the snippet presented in Listing~\ref{code:simple-expr}. 
}
\label{fig:graph-repr}
\end{subfigure}
\caption{AST and our graph representation for the small code snippet presented in Listing~\ref{code:simple-expr}.}
\label{fig:AST-Graph-Repr}
\end{figure*}

\section{Program Representations}
\label{sec:prog-repr}

We represent programs as directed graphs so the information can propagate in both directions in the \GNN. These graphs are based on the programs' \emph{abstract syntax trees} (\ASTs). An \AST is described by a set of nodes that correspond to non-terminal symbols in the programming language's grammar and a set of tokens that correspond to terminal symbols~\cite{hopcroft2008introduction}.
An \AST depicts a program's grammatical structure~\cite{compilers-book-dragon}. Figure~\ref{fig:AST-Graph-Repr}\subref{fig:AST} 
shows the \AST for the small code snippet presented in Listing~\ref{code:simple-expr}.

Regarding our graph program representation, firstly, we create a unique node in the \AST for each distinct variable in the program and connect all the variable occurrences in the program to the same unique node. Figure~\ref{fig:AST-Graph-Repr}\subref{fig:graph-repr} shows our graph representation for the small code snippet presented in Listing~\ref{code:simple-expr}. Observe that our representation uses a single node for each variable in the program, the green nodes \texttt{a} and \texttt{b}. Moreover, we consider five types of edges in our representation: child, sibling, read, write, and chronological edges. \emph{Child edges} correspond to the typical edges in the \AST representation that connect each parent node to its children. Child edges are bidirectional in our representation. In Figure~\ref{fig:AST-Graph-Repr}\subref{fig:graph-repr}, the black edges correspond to child edges. \emph{Sibling edges} connect each child to its sibling successor. These edges denote the order of the arguments for a given node and have been used in other program representations~\cite{iclr18-learning-to-represent-progs-as-graphs}. Sibling edges allow the program representation to differentiate between different arguments when the order of the arguments is important (e.g.\ binary operation such as $\le$). For example, consider the node that corresponds to the operation $\sigma(A_1, A_2, \ldots, A_m)$. The parent node $\sigma$ is connected to each one of its children by a child edge e.g.\ $\sigma \leftrightarrow A_1,\sigma \leftrightarrow A_2,\ldots,\sigma \leftrightarrow A_m$. Additionally, each child its connected to its successor by a sibling edge e.g.\ $A_1 \rightarrow A_2, A_2 \rightarrow A_3,\ldots, A_{m-1} \rightarrow A_m$. In Figure~\ref{fig:AST-Graph-Repr}\subref{fig:graph-repr}, the red dashed edges correspond to sibling edges.

Regarding the \emph{write and read edges}, these edges connect the \texttt{ID} nodes with the unique nodes corresponding to some variable. Write edges are connections between an \texttt{ID} node and its variable node. This edge indicates that the variable is being written.  
Read edges are also connections between an \texttt{ID} node and its variable node, although these edges indicate that the variable is being read.
In Figure~\ref{fig:AST-Graph-Repr}\subref{fig:graph-repr}, the blue dashed edge corresponds to a write edge while the green dashed edges correspond to read edges.
Lastly, \emph{chronological edges} establish an order between all the \texttt{ID} nodes connected to some variable. These edges denote the order of the \texttt{ID} nodes for a given variable node. For example, in Figure~\ref{fig:AST-Graph-Repr}\subref{fig:graph-repr}, the yellow dashed edge corresponds to a chronological edge between the \texttt{ID} nodes of the variable \texttt{a}.
Besides the siblings and the chronological edges, all the other edges are bidirectional in our representation.

\emph{The novelty of our graph representation} is that we create a unique variable node for each variable in the program and connect each variable's occurrence to its unique node. This lets us map two variables in two programs, even if their number of occurrences is different in each program. Furthermore, the variable's identifier is suppressed after we connect all the variable's occurrences to its unique node. This way, all the variables' identifiers are anonymized.
Prior work on representing programs as graphs~\cite{iclr18-learning-to-represent-progs-as-graphs,iclr19-learning-localize-and-repair,neurips21-bug-detection-repair} use different nodes for each variable occurrence and take into consideration the variable identifier in the program representation. Furthermore, to the best of our knowledge, combining all five types of edges (sibling, write, read, chronological, and \AST) is also novel. Section~\ref{sec:results-eval} presents an ablation study on the set of edges to analyze the impact of each type of edge.

\section{Graph Neural Networks (\GNNs)}
\label{sec:gnns}

Graph Neural Networks (\GNNs) are a subclass of neural networks designed to operate on graph-structured data \cite{DBLP:conf/iclr/KipfW17}, which may be citation networks~\cite{citnet}, mathematical logic \cite{DBLP:conf/itp/GoertzelJKOPU22} or representations of computer code~\cite{iclr18-learning-to-represent-progs-as-graphs}. Here, we use graph representations of a pair of \ASTs, representing two programs for which we want to match variables, as the input. The main operative mechanism is to perform \textit{message passing} between the nodes, so that information about the global problem can be passed between the local constituents. The content of these messages and the final representation of the nodes is parameterized by neural network operations (matrix multiplications composed with a non-linear function). For the variable matching task, we do the following to train the parameters of the network. After several message passing rounds through the edges defined by the program representations above, we obtain numerical vectors corresponding to each variable node in the two programs. We compute scalar products between each possible combination of variable nodes in the two programs, followed by a softmax function. 
Since the program samples are obtained by program mutation, the correct mapping of variables is known. Hence, we can compute a cross-entropy loss and minimize it so that the network output corresponds to the labeled variable matching. 
Note that the network has no information on the name of any object, which means that the task must be solved purely based on the structure of the graph representation. Therefore, our method is invariant to the consistent renaming of variables.

\paragraph{Architecture Details.}
The specific \GNN architecture used in this work is the relational graph convolutional neural network (RGCN), which can handle multiple edges or relation types within one graph~\cite{DBLP:conf/esws/SchlichtkrullKB18}.
The numerical representation of nodes in the graph is updated in the message passing step according to the following equation: 

\begin{align*}
    \mathbf{x}^{\prime}_i = \mathbf{\Theta}_{\textrm{root}} \cdot
        \mathbf{x}_i + \sum_{r \in \mathcal{R}} \sum_{j \in \mathcal{N}_r(i)}
        \frac{1}{|\mathcal{N}_r(i)|} \mathbf{\Theta}_r \cdot \mathbf{x}_j,
\end{align*}
where $\mathbf{\Theta}$ are the trainable parameters, $\mathcal{R}$ stands for the different edge types that occur in the graph, and $\mathcal{N}_r$ the neighbouring nodes of the current node $i$ that are connected with the edge type $r$~\cite{torchg}.
After each step, we apply Layer Normalization~\cite{DBLP:journals/corr/BaKH16} followed by a Rectified Linear Unit~(ReLU) non-linear function. 

We use two separate sets of parameters for the message passing phase for the program with the bug and the correct program. Five message passing steps are used in this work. After the message passing phase, we obtain numerical vectors representing every node in both graphs. We then calculate dot products $\vec{a} \cdot \vec{b}$ between the vectors representing variable nodes in the buggy program graph $a \in A$ and the variable nodes from the correct graph $b \in B$, where $A$ and $B$ are the sets of variable node vectors. A score matrix $\mathcal{S}$ with dimensions $|A| \times |B|$ is obtained, to which we apply the softmax function on each row to obtain the matrix $\mathcal{P}$. The values in each row of $\mathcal{P}$ can now be interpreted as representing the probability that variable $a_i$ maps to each of the variables $b_i$. 

\section{Use-Cases: Program Repair}
\label{sec:use-cases}

In this section, we propose a few use-cases on how to use variable mappings for program repair. More specifically, to repair bugs of: \textit{wrong comparison operator}, \textit{variable misuse}, and \textit{missing expression}. These bugs are common among novice programmers~\cite{codeFlaws-dataset} and have been studied by prior work in the field of automated program repair~\cite{iclr18-learning-to-represent-progs-as-graphs,oopsla18-deepBugs,iclr19-learning-localize-and-repair,neurips21-bug-detection-repair}.
The current state-of-the-art on semantic program repair tools focused on repairing \IPAs, such as \Clara~\cite{clara} and \Verifix~\cite{verifix}, are only able to fix these bugs if the correct expression in the correct program is located in a similar program structure as the incorrect expression in the incorrect implementation. For example, consider again the two programs presented in Figure~\ref{fig:progs-motivation}. If the loop condition was incorrect in the faulty program, \Clara and \Verifix could not fix it, since the control flow graphs do not match. Thus, these tools would fail due to \emph{structural mismatch}.

The following sections present three program repair tasks that take advantage of variable mappings to repair an incorrect program using a correct implementation for the same \IPA without considering the programs' structures. 
Our main goal is to show the usefulness of variable mappings.
We claim that variable mappings are informative enough to repair these three realistic types of bugs.
Given a buggy program, we search for and try to repair all three types of bugs.
Whenever we find a possible fix, we check if the program is correct using the \IPA's test suite.

\paragraph{\textbf{Bug \#1: Wrong Comparison Operator (WCO).}}
Our first use-case are faulty programs with the bug of wrong comparison operator (WCO). This is a recurrent bug in students' submissions to \IPAs since novice programmers frequently use the wrong operator, \break e.g., \texttt{i <= n} instead of \texttt{i < n}.

We propose tackling this problem solely based on the variable mapping between the faulty and correct programs, ignoring the programs' structure. First, we rename all the variables in the incorrect program based on the variable mapping by changing all the variables' identifiers in the incorrect program with the corresponding variables' identifiers in the correct implementation. Second, we count the number of times each comparison operation appears with a specific pair of variables/expressions in each program. Then, for each comparison operation in the correct program, we compute the mirrored expression, i.e., swapping the operator by its mirrored operator, and swapping the left-side and right-side of the operation. This way, if the incorrect program has the same correct mirrored expression, we can match it with an expression in the correct program. For example, in the programs shown in Figure~\ref{fig:progs-motivation}, both loop conditions would match even if they are mirrored expressions, i.e., \texttt{i <= n} and \texttt{n >= i}.

Afterwards, we iterate over all the pairs of variables/expressions that appear in comparison operations of the correct program (plus the mirrored expressions) and compare if the same pair of variables/expressions appear the same number of times in the incorrect program, using the same comparison operator. If this is not the case, we try to fix the program using the correct implementation's operator in each operation of the incorrect program with the same pair of variables/expressions. Once the program is fixed, we rename all the variables based on the reverse variable mapping.

\paragraph{\textbf{Bug \#2: Variable Misuse (VM).}}
Our second program repair task are buggy programs with variables being misused, i.e., the student uses the wrong variable in some program location. The wrong variable is of the same type as the correct variable that should be used. Hence, this bug does not produce any compilation errors. This type of bug is common among students and experienced programmers~\cite{dataset-ManySStuBs4J,icse20-graph2Diff-NNs}. 
The task of detecting this specific bug has received much attention from the Machine Learning (ML) research community~\cite{iclr18-learning-to-represent-progs-as-graphs,iclr19-learning-localize-and-repair,iclr21-source-code-representation}. 

Once again, we propose to tackle this problem based on the variable mapping between the faulty program and the correct one, ignoring the programs' structure. We start by renaming all the variables in the incorrect program based on the variable mapping. Then we count the number of times each variable appears in both programs. If a variable, \texttt{x}, appears more times in the incorrect program than in the correct implementation, and if another variable \texttt{y} appears more times in the correct program, then we try to replace each occurrence of \texttt{x} in the incorrect program with \texttt{y}. Once the program is fixed, we rename all the variables based on the reverse variable mapping.

\paragraph{\textbf{Bug \#3: Missing Expression (ME).}}
The last use-case we will focus on is to repair the bug of \emph{missing expressions/assignments}. This bug is also recurrent in students' implementations of \IPAs~\cite{codeFlaws-dataset}. Frequently, students forget to initialize some variable or to increment a variable of some loop, resulting in a bug of missing expression. However, unlike the previously mentioned bugs, this one has not received much attention from the ML community since it is more complex to repair this program fault. 
To search for a possible fix, we start by renaming all the variables in the incorrect program based on the variable mapping. Next, we count the number of times each expression appears in both programs. Expressions that appear more frequently in the correct implementation are considered possible repairs. 
Then, we try to inject these expressions, one at a time, into the incorrect implementation's code blocks and check the program's correctness.
Once the program is fixed, we rename all the variables based on the reverse variable mapping. This task is solely based on the variable mapping between the faulty and the correct programs.

\begin{table*}[t!]
\centering
\caption{Validation mappings fully correct after 20 training epochs.}
\scalebox{1}{\begin{tabular}{lllll}
\toprule
{} & \multicolumn{4}{c}{\textbf{Buggy Programs}} \\ 
\cline{2-5}
{}             & 
WCO Bug 
&
VM Bug 
&
ME Bug 
&
All Bugs 
\\ \toprule
Accuracy & \multicolumn{1}{r}{93.7\%} & \multicolumn{1}{r}{95.8\%} & \multicolumn{1}{r}{93.4\%} & \multicolumn{1}{r}{96.49\%} \\ \hline
\end{tabular}}
\label{tab:validation_perfs}
\end{table*}

\begin{table*}[t!]
\centering
\caption{
The number of correct variable mappings generated by our \GNN on the evaluation dataset and the average overlap coefficients between the real mappings and our \GNN's variable mappings. 
}
\scalebox{1}{%
\begin{tabular}{lllll}
\toprule
{} & \multicolumn{4}{c}{\textbf{Buggy Programs}} \\ 
 \cline{2-5}
\textbf{Evaluation Metric}                       & 
WCO Bug 
&
VM Bug 
&
ME Bug 
&
All Bugs 
\\ 
\toprule
  \# Correct Mappings & \multicolumn{1}{c}{87.38\%} & \multicolumn{1}{c}{81.87\%} & \multicolumn{1}{c}{79.95\%} & \multicolumn{1}{c}{82.77\%} \\ 
Avg Overlap Coefficient & \multicolumn{1}{c}{96.99\%} & \multicolumn{1}{c}{94.28\%} & \multicolumn{1}{c}{94.51\%} & \multicolumn{1}{c}{95.05\%} \\
\toprule
\# Programs & \multicolumn{1}{c}{1078} & \multicolumn{1}{c}{1936} & \multicolumn{1}{c}{1152} & \multicolumn{1}{c}{4166} \\ \hline

\end{tabular}

}
\label{tab:eval-metrics}
\end{table*}

\section{Experiments}
\label{sec:results}

\paragraph{\textbf{{Experimental Setup}.}}
We trained the Graph Neural Networks on an Intel(R) Xeon(R) Gold 6140 CPU @ 2.30GHz server with 72 CPUs and 692GB RAM. Networks were trained using NVIDIA GEFORCE GTX 1080 graphics cards with 12GB of memory.
All the experiments related to our program repair tasks were conducted on an Intel(R) Xeon(R) Silver computer with 4210R CPUs @ 2.40GHz, using a memory limit of 64GB and a timeout of 60 seconds.

\subsection{\IPAs Dataset}

To evaluate our work, we used \benchmark~\cite{C-Pack-IPAs}, a benchmark of student programs developed during an introductory programming course in the C programming language for ten different \IPAs, over two distinct academic years, at Instituto Superior Técnico.
These \IPAs are small imperative programs that deal with integers and input-output operations (see Appendix~\ref{appdx:IPAs-description}).

First, we selected a set of correct submissions, i.e., programs that compiled without any error and satisfied a set of input-output test cases for each \IPA. 
We gathered 238 correct students' submissions from the first year and 78 submissions from the second year.
We used the students' submissions from the first year for training and for validating our \GNN 
and the submissions from the second year for evaluating our work. 

Since we need to know the real variable mappings between programs (ground truth) to evaluate our representation, we generated a dataset of pairs of correct/incorrect programs to train and evaluate our work with specific bugs. This is a common procedure to evaluate machine learning models in the field of program repair~\cite{iclr18-learning-to-represent-progs-as-graphs,iclr19-learning-localize-and-repair,neurips21-bug-detection-repair,iclr21-source-code-representation, aitp22-learning-2-map-variables}. 
To generate this dataset, we used \multIPAs~\cite{fse22-MultIPAS}, a program modifier capable of mutating C programs syntactically, generating semantically equivalent programs, i.e., changing the program's structure but keeping its semantics.
There are several program mutations available in \multIPAs: mirroring comparison expressions, swapping the if's then-block with the else-block and negating the test condition, increment/decrement operators mirroring, variable declarations reordering, translating for-loops into equivalent while-loops, and all possible combinations of these program mutations. Hence, \multIPAs has thirty-one different configurations for mutating a program. All these program mutations generate semantically equivalent programs. 
Afterwards, we also used \multIPAs, to introduce bugs into the programs, such as 
\emph{wrong comparison operator} (WCO), \emph{variable misuse} (VM), \emph{missing expression} (ME).
Hence, we gathered a dataset of pairs of programs and the mappings between their sets of variables (see Appendix~\ref{appendix-dataset-gen}).
Each pair corresponds to a real correct student's implementation, and the second program is the student's program after being mutated and with some bug introduced.
Thus, this \IPA dataset is generated, although based on real programs. The dataset is divided into three different sets: training set, validation set, and evaluation set. The programs generated from \textit{first year} submissions are divided into a training and validation set based on which students' submissions they derive from. 80\% of the students supply the training data, while 20\% supply validation data. The evaluation set, which is not used during the machine learning optimization, is chronologically separate: it consists only of \textit{second year} submissions, to simulate the real-world scenario of new, incoming students. The training set is composed of 3372, 5170, and 2908  pairs of programs from the first academic year for the 
WCO, VM, and ME bugs, respectively. The validation set, which was used during development to check the generalization of the prediction to unseen data, comprises 1457, 1457, and 1023  pairs of programs from the first year. 
Note that we subsample from the full spectrum of possible mutations, to keep the training data size small enough to train the network with reasonable time constraints. From each of the 31 combinations of mutations, we use one randomly created sample for each student per exercise. We found that this already introduced enough variation in the training dataset to generalize to unseen data.
Finally, the evaluation set is composed of 4166 pairs of programs from the second year (see $3^{rd}$ row, Table~\ref{tab:eval-metrics}).
This dataset will be publicly available for reproducibility reasons.

\subsection{Training}

At training time, since the incorrect program is generated, the mapping between the variables of both programs is known. The network is trained by minimizing the cross entropy loss between the labels (which are categorical integer values indicating the correct mapping) and the values in each corresponding row of the matrix $\mathcal{P}$. As an optimizer, we used the Adam algorithm with its default settings in PyTorch~\cite{DBLP:journals/corr/KingmaB14}. The batch size was 1. As there are many different programs generated by the mutation procedures, we took one sample from each mutation for each student. Each network was trained for 20 full passes (epochs) over this dataset while shuffling the order of the training data before each pass. For validation purposes, data corresponding to 20$\%$ of the students from the first year of the dataset was kept separate and not trained on.

Table~\ref{tab:validation_perfs} shows the percentage of validation data mappings that were exactly correct (accuracy) after 20 epochs of training, using four different \GNN models. Each \GNN model was trained on programs with the bugs of wrong comparison operator (WCO), variable misuse (VM), missing expression (ME) or all of them (All). Furthermore, each \GNN model has its own validation set with programs with a specific type of bug. The \GNN model trained on All Bugs was validated using a mix of problems from each bug type. In the following sections, we focus only on this last \GNN model (All Bugs).

\begin{table*}[t!]
    \centering
    \caption{Percentage of variable mappings fully correct on the validation set for different sets of edges used. Each type of edge is represented by an index using the mapping: \{0: \AST; 1: sibling; 2: write; 3: read; 4: chronological\}. }
    \begin{tabular}{lrrrrrrr}
\toprule
\textbf{Edges Used} &   
All &
(1,2,3,4) &
(0,2,3,4) &
(0,1,3,4) &
(0,1,2,4) &
(0,1,2,3) &
(0,1) \\
\midrule
\textbf{Accuracy} &
\textbf{96.49\% }&
52.53\% &
73.76\% &
95.45\% &
94.87\% &
96.06\% &
94.74\%\\
\bottomrule
\end{tabular}
\label{tab:ablation-study}
\end{table*}

\subsection{Evaluation}
\label{sec:results-eval}

Our \GNN model was trained on programs with bugs of wrong comparison operator (WCO), variable misuse (VM), and missing expression (ME).
We used two evaluation metrics to evaluate the variable mappings produced by the \GNN. 
First, we counted the number of totally correct mappings our \GNN was able to generate. We consider a variable mapping totally correct if it correctly maps all the variables between two programs. Secondly,
we computed the overlap coefficient between the original variable mappings and the variable mappings generated by our \GNN. The overlap coefficient is a similarity metric given by the intersection between the two mappings divided by the length of the variable mapping (see Appendix~\ref{appendix-overlap-coeff}).

The first row in Table~\ref{tab:eval-metrics} shows the number of totally correct variable mappings computed by our \GNN model. One can see that the \GNN maps correctly around 83\% of the evaluation dataset.
We have also looked into the number of variables in the mappings we were not getting entirely correct. The results showed that programs with more variables  (e.g., six or seven variables) are the most difficult for our \GNN to map their variables correctly (see Appendix~\ref{appendix-num-vars}).
For this reason, we have also computed the overlap coefficient between the \GNN's variables mappings and the original mappings (ground truth).
The second row in Table~\ref{tab:eval-metrics} shows the average of the overlap coefficients between the original variable mappings and the mappings generated by our \GNN model.
The overlap coefficient~\cite{mlap16-survey-similarity-measures} measures the intersection (overlap) between two mappings. If the coefficient is $100\%$, both sets are equal. One set cannot be a subset of the other since both sets have the same number of variables in our case. The opposite is $0\%$ overlap, meaning there is no intersection between the two mappings.
The \GNN achieved at least 94\% of overlap coefficients, i.e., even if the mappings are not always fully correct, almost 94\% of the variables are correctly mapped by the \GNN. 

\begin{table*}[t!]
\centering
\caption{The number of programs repaired by each different repair technique: 
\Verifix, \Clara, and our repair approach based on our \GNN's variable mappings. 
The first row shows the results of repairing the programs using variable mappings generated based on uniform distributions (baseline).
}
\resizebox{\textwidth}{!}{%
\begin{tabular}{@{\extracolsep{8pt}}ccccccc}

\toprule
{} & \multicolumn{4}{c}{\textbf{Buggy Programs}} & \multicolumn{2}{c}{\textbf{Not Succeeded}} \\ \cline{2-5} \cline{6-7}
\textbf{Repair Method} &
WCO Bug 
&
VM Bug 
&
ME Bug 
&
All Bugs 
&
\textbf{\% Failed} &
\textbf{\% Timeouts (60s)}
\\ 
\toprule

\textbf{Baseline} & 618 (57.33\%) & 1187 (61.31\%) & 287 (24.91\%) & 2092 (50.22\%) & 0 (0.0\%) & \textbf{2074 (49.78\%)}\\
\textbf{\Verifix} & 555 (51.48\%) & 1292 (66.74\%) & 741 (64.32\%) & 2588 (62.12\%) & \textbf{1471 (35.31\%)} & 107 (2.57\%)\\
\textbf{\Clara} & 722 (66.98\%) & 1517 (78.36\%) & 764 (66.32\%) & 3003 (72.08\%) & 1153 (27.68\%) & 10 (0.24\%)\\
\textbf{GNN} & \textbf{992 (92.02\%)} & \textbf{1714 (88.53\%)} & \textbf{981 (85.16\%)} & \textbf{3687 (88.5\%)} & 0 (0.0\%) & 479 (11.5\%)\\
  \bottomrule
\end{tabular}%
}

\label{tab:program-repair-results}
\end{table*}

\paragraph{\textbf{Ablation Study.}} To study the effect of each type of edge in our program representation, we have performed an ablation study on the set of edges. Prior works have done similar ablation studies~\cite{iclr18-learning-to-represent-progs-as-graphs}.
Table~\ref{tab:ablation-study} presents the accuracy of our \GNN (i.e., number of correct mappings) on the evaluation dataset after 20 epochs. We can see that the accuracy of our \GNN drops from 96\% to 53\% if we remove the AST edges (index 0), which was expected since these edges provide syntactic information about the program. Removing the sibling edges (index 1) also causes a great impact on the \GNN's performance, dropping to 74\%. The other edges are also important, and if we remove them, there is a negative impact on the \GNN's performance. Lastly, since the AST and sibling edges caused the greatest impact, we evaluated using only these edges on our \GNN and got an accuracy of 94.7\%. However, the model using all the proposed edges has the highest accuracy of 96.49\%.

\subsection{Program Repair}
\label{sec:results-prog-repair}

This section presents the results of using variable mappings on the three use-cases described in Section~\ref{sec:use-cases}, i.e., the tasks of repairing bugs of: \textit{wrong comparison operator} (WCO), \textit{variable misuse} (VM) and \textit{missing expression} (ME). For this evaluation, we have also used the two current publicly available 
program repair tools for fixing introductory programming assignments (\IPAs): \Clara~\cite{clara} and \Verifix~\cite{verifix}. 
Furthermore, we have tried to fix each pair of incorrect/correct programs in the evaluation dataset by passing each one of these pairs of programs to every repair method: \Verifix, \Clara, and our repair approach based on the \GNN's variable mappings. 

If our repair procedure cannot fix the incorrect program using the most likely variable mapping according to the \GNN model, then it generates the next most likely mapping based on the variables' distributions computed by the \GNN. Therefore, the repair method iterates over all variable mappings based on the \GNN's predictions.
Lastly, we have also run the repair approach using as baseline variable mappings generated based on uniform distributions. This case simulates most repair techniques that compute all possible mappings between both programs' variables (e.g., \textsc{SearchRepair}~\cite{searchRepair}).

Table~\ref{tab:program-repair-results} presents the number of programs repaired by each different repair method. 
The first row presents the results for the baseline, which was only able to fix around 50\% of the evaluation dataset.
In the second row, the interested reader can see that \Verifix can only repair about 62\% of all programs. \Clara, presented in the third row, outperforms \Verifix, 
being able to repair around 72\% of the whole dataset. 
The last row presents the \GNN model. This model is the best one repairing 88.5\% of the dataset.

The number of executions that resulted in a timeout (60 seconds) is relatively small for \Verifix and \Clara. Regarding our repair procedure, it either fixes the incorrect program or iterates over all variable mappings until it finds one that fixes the program. Thus, the baseline and the \GNN present no failed executions and considerably high rates of executions that end up in timeouts, almost 50\% for the baseline and 11.5\% in the case of the \GNN model.
Additionally, Table~\ref{tab:program-repair-results} also presents the failure rate of each technique, i.e., all the computations that ended within 60 seconds and did not succeed in fixing the given incorrect program. \Verifix has the highest failure rate, around 35\% of the entire evaluation set. \Clara also presents a significant failure rate, about 28\%. As explained previously, this is the main drawback of these tools.
Hence, these results support our claim that it is possible to repair these three realistic bugs solely based on the variable mappings' information without matching the structure of the incorrect/correct programs.

Furthermore, considering all executions, the average number of variable mappings used within 60 seconds is 1.24 variable mappings for the \GNN model and 5.6 variable mappings when considering the baseline. The minimum number of mappings generated by both approaches is 1, i.e., both techniques were able to fix at least one incorrect program using the first generated variable mapping. The maximum number of variable mappings generated was 32 (resp. 48) for the \GNN (resp. baseline). The maximum number of variable mappings used is high because the repair procedure iterates over all the variable mappings until the program is fixed or the time runs out. Moreover, even if we would only consider using the first variable mapping generated by the \GNN model to repair the incorrect programs, we would be able to fix 3377 programs in 60 seconds, corresponding to 81\% of the evaluation dataset.

\begin{figure}[t!]
    \centering
    \resizebox{0.45\textwidth}{!}{\includegraphics{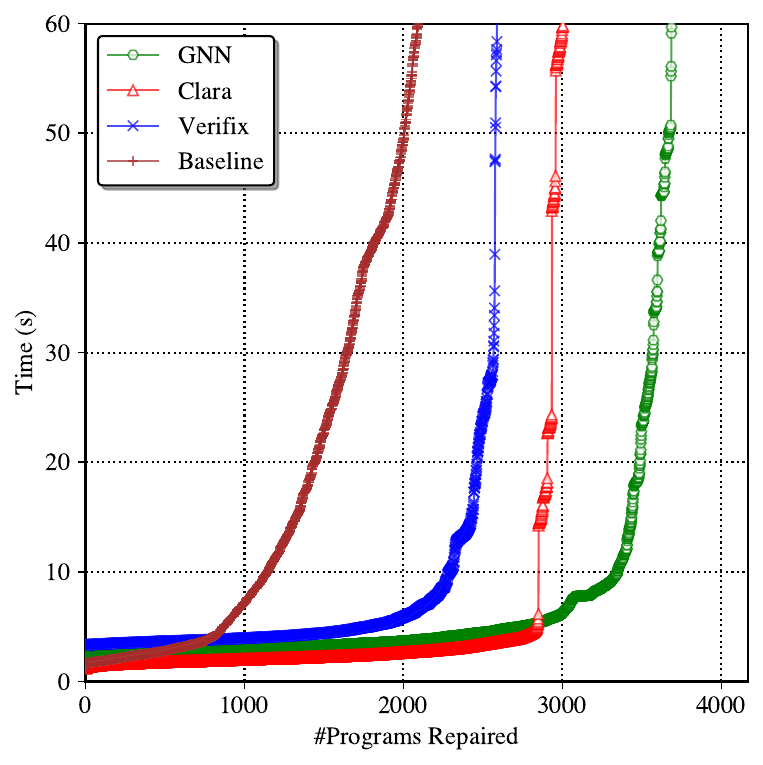}}
    \caption{Cactus plot - The time spent by each method repairing each program of the evaluation dataset, using a timeout of 60 seconds.}
    \label{fig:cactus-time}
\end{figure}

Regarding the time performance of each technique, Figure~\ref{fig:cactus-time} shows a cactus plot that presents the CPU time spent, in seconds, on repairing each program ($y$-axis) against the number of repaired programs ($x$-axis) using different repairing techniques. One can clearly see a gap between the different repair methods' time performances. For example, in 10 seconds, the baseline can only repair around 1150 programs, \Verifix repairs around  2300, \Clara repairs around 2850 programs while using the \GNN's variable mappings, we can repair around 3350 programs, i.e., around 17\% more. 
We are considering the time the \GNN takes to generate the variable mappings and the time spent on the repair procedure. However, the time spent by the \GNN to generate one variable mapping is almost insignificant. The average time the \GNN takes to produce a variable mapping is 0.025 seconds. The minimum (resp. maximum) time spent by the \GNN, considering all the executions is 0.015s (resp. 0.183s).

\section{Related Work}
\label{sec:related-work}

\emph{Automated program repair}~\cite{verifix,autograder,clara,deepfix,drRepair} has become crucial to provide feedback to novice programmers by checking their introductory programming assignments (\IPAs) submissions using a test suite.
In order to repair an incorrect program with a correct reference implementation, \Clara~\cite{clara} requires a perfect match between both programs' control flow graphs and a bijective relation between both programs' variables. Otherwise, \Clara returns a structural mismatch error.
\Verifix~\cite{verifix} aligns the control flow graph (CFG) of an incorrect program with the reference solution's CFG. Then, using that alignment relation and \textsc{MaxSMT} solving, \Verifix proposes fixes to the incorrect program. \Verifix also requires a compatible control flow graph between the incorrect and the correct program.
\textsc{BugLab}~\cite{neurips21-bug-detection-repair} is a Python program repair tool that learns how to detect and fix minor semantic bugs. To train \textsc{BugLab}, \cite{neurips21-bug-detection-repair} applied four program mutations and introduced four different bugs to augment their benchmark of Python programs.
\textsc{DeepBugs}~\cite{oopsla18-deepBugs} uses rule-based mutations to build a dataset of programs from scratch to train its ML-based program repair tool. 
Given a program, this tool classifies if the program is buggy or not.

\emph{Mapping variables} can also be helpful for the task of \emph{code adaption}, where the repair framework tries to adapt all the variable names in a pasted snippet of code, copied from another program or a Stack Overflow post to the surrounding preexisting code~\cite{corr22-adaptivePaste}.
\textsc{AdaptivePaste}~\cite{corr22-adaptivePaste} focused on a similar task to \emph{variable misuse} (VM) repair, it uses a sequence-to-sequence with multi-decoder transformer training to learn programming language semantics to adapt variables in the pasted snippet of code.
Recently, several systems were proposed to tackle the VM bug with ML models~\cite{iclr18-learning-to-represent-progs-as-graphs,iclr20-global-relational-source-code,oopsla20-learning-prog-embeddings-ginn}.
These tools classify the variable locations as faulty or correct and then replace the faulty ones through an enumerative prediction of each buggy location~\cite{iclr18-learning-to-represent-progs-as-graphs}. However, none of these methods takes program semantics into account, especially the long-range dependencies of variable usages~\cite{corr22-adaptivePaste}.

\section{Conclusions}
\label{sec:conclusions}

This paper tackles the highly challenging problem of mapping variables between programs.
We propose the usage of graph neural networks (\GNNs) to map the set of variables between two programs using our novel graph representation that is based on both programs' abstract syntax trees.
In a dataset of 4166 pairs of incorrect/correct programs, experiments show that our \GNN correctly maps 83\% of the evaluation dataset. 
Furthermore, we leverage the variable mappings to perform automatic program repair. While the current state-of-the-art on program repair can only repair about 72\% of the evaluation dataset due to structural mismatch errors, our approach, based on variable mappings, is able to fix 88.5\%.

In future work, we propose to integrate our variable mappings into other program repair tools to evaluate the impact of using these mappings to repair other types of bugs. Additionally, we will analyze using our mappings to fix an incorrect program using several correct programs.

\section*{Acknowledgements}

This work was supported by Portuguese national funds through FCT under projects UIDB/50021/2020, PTDC/CCI-COM/2156/2021, 2022.03537.PTDC and grant SFRH/\-BD/\-07724/\-2020. This work was also supported by European funds through COST Action CA2011; by the European Regional Development Fund under the Czech project AI\&Reasoning no.~CZ.02.1.01/0.0/0.0/15\_003/0000466 (JP), Amazon Research Awards (JP), and by the Ministry of Education, Youth, and Sports within the program ERC CZ under the project POSTMAN no. LL1902. This article is part of the RICAIP project that has received funding from the EU’s Horizon 2020 research and innovation program under grant agreement No 857306.

\appendix

\begin{figure*}[t!]
    \centering
    \resizebox{\textwidth}{!}{\includegraphics{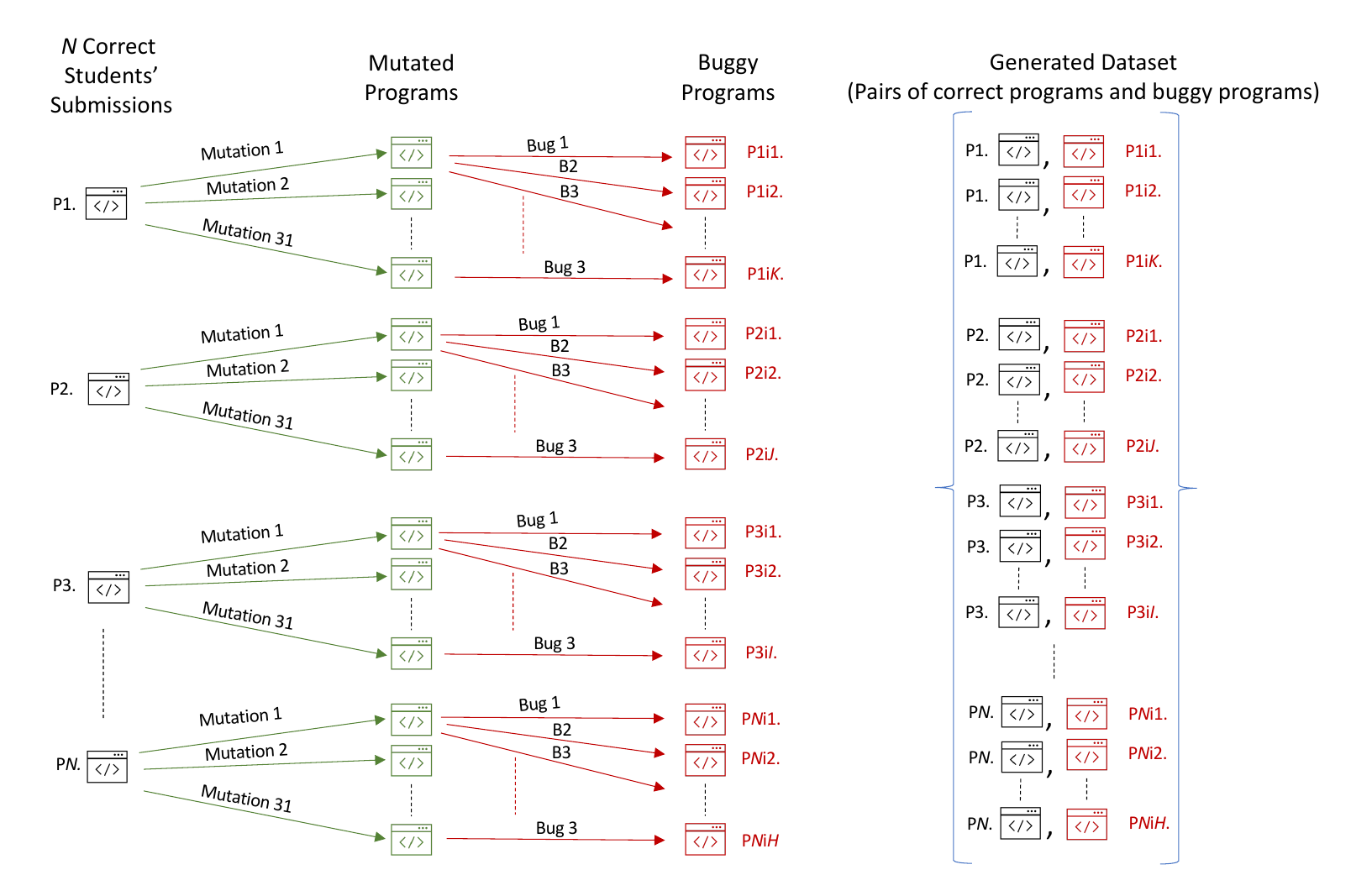}}
    \caption{\IPAs Dataset Generation}
    \label{fig:dataset-generation}
\end{figure*}

\section{\IPAs Dataset Generation}
\label{appendix-dataset-gen}
To evaluate our work, we have generated a dataset of pairs programs based on a benchmark of student programs developed during an introductory programming course in the C programming language for ten different introductory programming assignments (\IPAs), over two distinct academic years. We selected only semantically correct submissions i.e., programs that compiled without any error and satisfied a set of input-output test cases for each \IPA. 

Afterwards, we generated a dataset of pairs of correct/incorrect programs to train and evaluate our work with specific bugs. The reason to generate programs is that we need to know the real variable mappings between two programs (ground truth) to evaluate our representation. 
As explained in the paper, we used \multIPAs~\cite{fse22-MultIPAS}
to generate this dataset.
This tool can mutate our programs syntactically, generating semantically equivalent programs.
There are several program mutations available in \multIPAs such as: mirroring comparison expressions, swapping the if's then-block with the else-block and negating the test condition, increment/decrement operators mirroring, variable declarations reordering, translating for-loops into equivalent while-loops, and all possible combinations of these program mutations. Hence, \multIPAs has 31 different configurations for mutating a program. Each program mutation can be applied in more than one place for a given program. Hence, each program mutation can generate several different mutated programs. For example, using the program mutation that reorders variable declarations, each possible reordering generates a different mutated program.

Regarding the generation of buggy programs, we also used \multIPAs, for introducing bugs into the programs, such as \textit{wrong comparison operator} (WCO), \textit{variable misuse} (VM) and \textit{missing expression} (ME). Each bug can be applied in more than one place for a given program. Thus, one program can generate several different buggy programs using the same bug. For example, the bug of variable misuse can be applied in each variable occurrence in the program, each one generates a single buggy program.

Figure~\ref{fig:dataset-generation} presents the generation of our dataset. 
Firstly, we applied all the available program mutations to each correct student's submission.
Then, for each mutated program, we applied all three types of bugs: WCO, VM and ME. 
Finally, we gathered a dataset of pairs of programs and the mappings between their sets of variables. As Figure~\ref{fig:dataset-generation} shows, each pair of programs, in our generated dataset, corresponds to a correct student's implementation and the student's program after being mutated and with some bug introduced.

\section{Description of \IPAs}
\label{appdx:IPAs-description}
The set of Introductory Programming Assignments (\IPAs) used to train and evaluate the \GNN model is part of the \benchmark benchmark~\cite{C-Pack-IPAs}.
In this set of \IPAs the students learn how to program with integers, floats, IO operations (mainly \texttt{printf} and \texttt{scanf}), conditionals (if-statements), and simple loops (for and while-loops).

\paragraph{\IPA\#1.} Write a program that determines and prints the largest of three integers given by the user.

\paragraph{\IPA\#2.} Write a program that reads two integers `N, M` and prints the smallest of them in the first row and the largest in the second.

\paragraph{\IPA\#3.} Write a program that reads two positive integers `N, M` and prints "yes" if `M` is a divisor of `N`, otherwise prints "no".

\paragraph{\IPA\#4.} Write a program that reads three integers and prints them in order on the same line. The smallest number must appear first.

\paragraph{\IPA\#5.} Write a program that reads a positive integer `N` and prints the numbers `1..N`, one per line.

\paragraph{\IPA\#6.}Write a program that determines the largest and smallest number of `N` real numbers given by the user. Consider that `N` is a value requested from the user.
The result must be printed with the command `printf("min: \%f, max: \%f\~n", min, max)`.

\paragraph{\IPA\#7.} Write a program that asks the user for a positive integer `N` and prints the number of divisors of `N`. Remember that prime numbers have 2 divisors.

\paragraph{\IPA\#8.}  Write a program that calculates and prints the average of `N` real numbers given by the user. The program should first ask the user for an integer `N`, representing the number of numbers to be entered. The real numbers must be represented
by float type. The result must be printed with the command `printf("\%.2f", avg);`.

\paragraph{\IPA\#9.} Write a program that asks the user for a value `N` corresponding to a certain period of time in seconds. The program should output this period of time in the format `HH:MM:SS`.

\paragraph{\IPA\#10.} Write a program that asks the user for a positive value `N`. The output should present the number of digits that make up `N` (on the first line), as well as the sum of the digits of `N` (on the second line). For example, the number 12345 has 5 digits, and the sum of these digits is 15.

\section{\#Correct/Incorrect Mappings vs \#Variables}
\label{appendix-num-vars}

\begin{figure*}
    \centering
    \resizebox{0.75\textwidth}{!}{\includegraphics{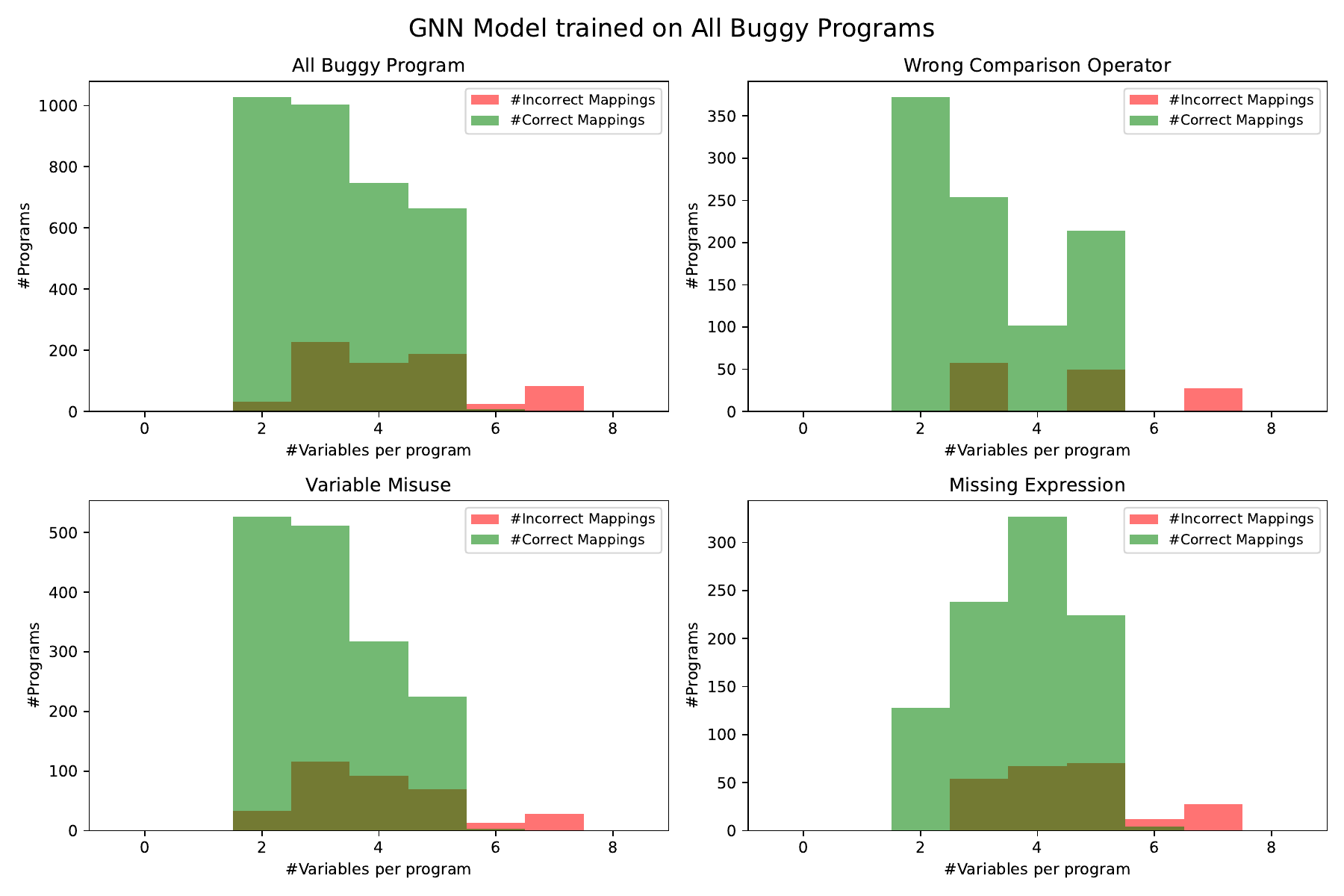}}
    \caption{Histograms showing the number of programs that our \GNN, trained on all buggy programs, mapped all their variables correctly. The results are presented for programs with the bugs of wrong comparison operator (WCO), variable misuse (VM), missing expression (ME) or all of them (All).}
    \label{fig:histogram-model-all}
\end{figure*}

Figure~\ref{fig:histogram-model-all} shows a histogram with the number of programs, $y$-axis, whose variables (number of variables in the $x$-axis) our \GNN models can map totally correct (\textcolor{green!70!black}{\#Correct Mappings}) in green and programs with at least one variable being mapped incorrectly (\textcolor{red!90!black}{\#Incorrect Mappings}) in red.

\section{Overlap Coefficient}
\label{appendix-overlap-coeff}

The overlap or Szymkiewicz–Simpson coefficient measures the overlap between two sets (e.g. mappings). This metric can be calculated by dividing the size of the intersection of two sets by the size of the smaller set, as follows:

\begin{equation}
    \label{eq:overlap}
    overlap(A,B) = \frac{| A \cap B |}{min(|A|,|B|)}
\end{equation}

An overlap of $100\%$ means that both sets are equal or one of them is a subset of the other. The opposite, $0\%$ overlap, means there is no intersection between both sets.

\bibliography{mybibliography}
\end{document}